\documentclass[12pt]{article}

\usepackage{amssymb,epsfig}

\begin{document}

\begin{center}
\baselineskip=24pt

{\Large \bf Limits on WIMP cross-sections from the 
NAIAD experiment at the Boulby Underground Laboratory}
\vspace{0.5cm}

\baselineskip=18pt

{\center \large \bf The UK Dark Matter Collaboration}

\vspace{0.5cm}

{\large G.~J.~Alner~$^a$, H.~M.~Ara\'ujo~$^b$, G.~J.~Arnison~$^a$,
J.~C.~Barton~$^c$~\footnote{deceased}, 
A.~Bewick~$^b$, C.~Bungau~$^b$, B.~Camanzi~$^b$,
M.~J.~Carson~$^d$, D.~Davidge~$^b$, E.~Daw~$^d$, 
J.~V.~Dawson~$^b$, G.~J.~Davies~$^b$,
J.~C.~Davies~$^d$, C.~Duffy~$^d$, T.~J.~Durkin~$^a$, T.~Gamble~$^d$,
S.~P.~Hart~$^a$, R.~Hollingworth~$^d$, G.~J.~Homer~$^a$, 
A.~S.~Howard~$^b$, I.~Ivaniouchenkov~$^{a,b}$, W.~G.~Jones~$^b$,
M.~K.~Joshi~$^b$, J.~Kirkpatrick~$^d$,
V.~A.~Kudryavtsev~$^d$~\footnote{Corresponding author, 
e-mail: v.kudryavtsev@sheffield.ac.uk}, T.~B.~Lawson~$^d$,
V.~Lebedenko~$^b$,
M.~J.~Lehner~$^d$~\footnote{Now at the University of Pennsylvania, 
Philadelphia, PA 19104, USA}, J.~D.~Lewin~$^a$, P.~K.~Lightfoot~$^d$, 
I.~Liubarsky~$^b$,
R.~L\"uscher~$^{a,b}$, J.~E.~McMillan~$^d$, 
B.~Morgan~$^d$,
A.~Murphy~$^e$, A.~Nickolls~$^a$, G.~Nicklin~$^d$,
S.~M.~Paling~$^d$, R.~M.~Preece~$^a$, J.~J.~Quenby~$^b$,
J.~W.~Roberts~$^{a,d}$,
M.~Robinson~$^d$, N.~J.~T.~Smith~$^a$, P.~F.~Smith~$^a$,
N.~J.~C.~Spooner~$^d$~\footnote{Corresponding author, 
e-mail: n.spooner@sheffield.ac.uk},
T.~J.~Sumner~$^b$, D.~R.~Tovey~$^d$, E.~Tziaferi~$^d$}

\vspace{0.5cm}
$^a$ {\it Particle Physics Department, 
Rutherford Appleton Laboratory, \\
OX11 0QX, UK}

$^b$ {\it Blackett Laboratory, Imperial College London, 
SW7 2BZ, UK}

$^c$ {\it Department of Physics, Queen Mary -- University of 
London, E1 4NS, UK}

$^d${\it Department of Physics and Astronomy, 
University of Sheffield, S3 7RH, UK}

$^e${\it School of Physics, 
University of Edinburgh, EH9 3JZ, UK}

\vspace{0.5cm}
\begin{abstract}
The NAIAD experiment (NaI Advanced Detector) for WIMP
dark matter searches at the Boulby Underground
Laboratory (North Yorkshire, UK) ran
from 2000 until 2003. A total of 44.9~kg$\times$years of
data collected with 2 encapsulated and 4 unencapsulated
NaI(Tl) crystals with high light yield were included in the analysis.
We present final results of this analysis carried
out using pulse shape discrimination.
No signal associated with nuclear recoils from WIMP interactions
was observed in any run with any crystal. 
This allowed us to set upper limits on the 
WIMP-nucleon spin-independent and WIMP-proton spin-dependent
cross-sections. The NAIAD experiment has so far imposed the most stringent
constraints on the spin-dependent WIMP-proton cross-section.
\end{abstract}

\end{center}

\vspace{0.5cm}
\noindent {\it Key words:} Dark matter, WIMP, neutralino,
Pulse shape analysis, Scintillation detectors, Inorganic crystals

\noindent {\it PACS:}  95.35.+d, 95.30.Cq, 14.80.Ly, 29.40.Mc

\vspace{0.5cm}
\noindent Corresponding author: V. A. Kudryavtsev, N. J. C. Spooner,
Department of Physics and Astronomy, University of Sheffield, 
Hicks Building, Hounsfield Rd., 
Sheffield S3 7RH, UK

\noindent Tel: +44 (0)114 2224531; \hspace{2cm} Fax: +44 (0)114 2728079; 

\noindent E-mail: v.kudryavtsev@sheffield.ac.uk, n.spooner@sheffield.ac.uk

\pagebreak

{\large \bf 1. Introduction}
\vspace{0.3cm}

\indent The UK Dark Matter Collaboration (UKDMC) has been operating 
various detectors for non-baryonic dark matter searches at the Boulby 
Underground Laboratory (North Yorkshire, UK) for many years. 
Limits on the flux of weakly interacting massive 
particles (WIMPs) -- the primary candidate for the non-baryonic dark 
matter that may 
constitute up to $90\%$ of the mass of the Galaxy, were set
using data from
the first NaI(Tl) detector \cite{ukdmc1,ukdmc3} and later improved
with an array of several NaI(Tl) crystals (NAIAD) \cite{naiad1}. 
Pulse shape analysis (PSA) was applied to the data
to distinguish between slow scintillations arising 
from background electron recoils and fast scintillations 
due to nuclear recoils, which are expected from WIMP-nucleus 
interactions \cite{dan1}. 

The advance of the UKDMC NaI(Tl) experiment was blocked
for a few years (1997-2000) by the discovery of a fast anomalous
component in the data from several encapsulated crystals
\cite{ukdmc2}. These events were faster than typical electron 
recoil pulses and faster even than nuclear recoil pulses 
\cite{ukdmc2,vak}. 
Similar events were also seen by
the Saclay group \cite{gerbier}.
These events were later shown to be due to implanted surface 
contamination of the crystal by an alpha-emitting
isotope from radon decay \cite{smith,csitest}. 

NAIAD was constructed in 2000-2001 with 5 unencapsulated crystals
to allow better control of the crystal surface, to reduce
the background due to surface events and to improve
the light collection \cite{naiad1,naiad}. In 2002 two more
encapsulated crystals were added to the array.

Despite the limited sensitivity of inorganic scintillators to WIMP
interactions, interest in the NaI(Tl) crystals has remained high 
for the past several years because of the DAMA group's claim 
of an annual modulation in their NaI(Tl) array consistent
with the expected signal from WIMP-nucleus interactions with
a specific set of WIMP parameters \cite{dama}.

NaI has the advantage of having two target nuclei with high and low
masses, thus reducing uncertainties related to nuclear
physics calculations. The detectors are sensitive to both
spin-independent and spin-dependent interactions. The NAIAD
experiment is complementary to other dark matter experiments
at Boulby, such as ZEPLIN \cite{zeplin} (liquid xenon and two phase
xenon detectors) and DRIFT \cite{drift} (time projection chamber
with directional sensitivity).
The array of NaI(Tl) detectors was also used as a
diagnostic array to study background and systematic effects
for other experiments at Boulby.

First results from the NAIAD experiment (10.6~kg$\times$years
exposure from 2000-2001) were published in Ref. \cite{naiad1}. 
Preliminary limits from an extended set of data (combined with
the previous set) with total exposure of 20.3~kg$\times$years
(2000-2002) were set in Ref. \cite{icrc}.
In this paper we present results from the NAIAD experiment
based on all data sets collected in 2000-2003.

\vspace{0.5cm}
{\large \bf 2. The NAIAD experiment}
\vspace{0.3cm}

The NAIAD array was sited in the underground laboratory
at Boulby mine at a vertical depth
of 1070 metres or 2805 m~w.~e. \cite{robinson}.
It consisted of 7
NaI(Tl) crystals from different manufacturers (Bicron, Hilger,
VIMS and Crismatec--Saint-Gobain) with a
total mass of about 55 kg. Two detectors contained encapsulated
crystals, while 5 other crystals were unencapsulated.
To avoid their degradation by humidity in the atmosphere, the
unencapsulated crystals were sealed in copper boxes filled
with dry nitrogen.

One of the crystals (DM70-Saclay) of similar design to those
used by the DAMA group in its 100~kg array, was previously running at
Modane and then moved to Boulby in 2001. When encapsulated, it showed
the anomalous fast population of events \cite{gerbier,csitest},
which were explained as being due to implanted surface 
contamination of the crystal by an alpha-emitting
isotope from radon decay \cite{smith,csitest}.
It was de-encapsulated in 2002, polished and was running as
part of the NAIAD array in 2002-2003. No fast population
was seen in the crystal after de-encapsulation \cite{matt,matt1}.

Each crystal was mounted in a 10 mm thick solid
PTFE (polytetrafluoroethylene) reflector cage and was coupled 
to light guides. The
two 4-5 cm long quartz light guides were also mounted in the 
PTFE cages and were coupled to 5 inch diameter
low background photomultiplier tubes (PMTs), ETL type 9390UKB.
Only selected low background materials were used in the detector design
including oxygen-free high-conductivity copper and PTFE.

Temperature control of the system was achieved through copper coils
outside the copper box. Chilled water was constantly pumped 
through the coils maintaining the temperature of the crystals 
stable to within a fraction of a degree during a single run.
The temperature of the crystal,
ambient air, water in the pipes and copper was measured
by thermocouples. If, for any
reason (for example, chiller failure), the variation of crystal 
temperature exceeded the predefined limit, the data from these periods
were not included in the analysis. If any changes in the experimental
set-up resulted in a change of the mean temperature of the crystal,
the data from different runs were not combined together, but assumed
to come from different experiments, so only the resulting limits
were combined.

Pulses from individual PMTs 
were integrated using a buffer circuit and then digitised using an
Acqiris CompactPCI based data acquisition (DAQ) system. 
Note that previous results
\cite{naiad1} were obtained with a digital oscilloscope and
Labview-based DAQ software.
The digitised pulse shapes (5 $\mu$s total digitisation time, 
10 ns digitisation accuracy) 
were passed to a computer running Linux OS and stored on disk. 
The gain of the PMTs was set to give about 2.5 mV per 
photoelectron. 
Low threshold discriminators were set to about 10 mV threshold,
which corresponded to about 1-1.5 keV of electron equivalent energy. 
A software threshold was
set at 4 keV since no pulse shape discrimination was observed below
this energy \cite{naiad1}.

Copper boxes containing the crystals were installed in 
lead and copper ``castles'', to shield the detectors from
background due to natural radioactivity in the surrounding rock. 
Wax and polypropylene neutron shielding (about 10~g/cm$^2$) 
around the castles
was installed in Spring 2002 and all data presented here 
(2002-2003) were collected with neutron shielding in place.

The light yield of the crystals was obtained from measurement of
the single photoelectron pulses 
and the energy calibration with a 122 keV $^{57}$Co gamma-ray source.
The crystals had light yields of 4.6-8.4 pe/keV. 
The large difference in the crystal light yield is not surprising
taking into account the different origin (manufacturers) of the
crystals, encapsulation, reflecting material,
size of the light guides and PMT response. The light yield was
checked every 2-4 weeks and was found to be stable within 10\% for
all crystals over long running periods (several months and more). 
The longest operated unencapsulated crystal, 
running since the summer of 2000, showed
a degradation in light yield of no more than 10\%.
Small variations in the light yield do not affect the energy
threshold for the data analysis since it is significantly higher
(4~keV) than the effective hardware threshold (typically less than 2~keV).
Possible short term variations in the light yield, as well as 
any electronic problems (stability of the discriminator thresholds,
high-voltage power supplies etc.) were monitored by plotting
the energy spectra of gamma calibration data and real data for each
day of running and removing suspicious data from the analysis. 
Any variations in the experimental conditions, mentioned above, 
would manifest themselves in the energy spectra. 

\vspace{0.5cm}
{\large \bf 3. Analysis procedure}
\vspace{0.3cm}

Final analysis was 
performed on the sum of the pulses from the two PMTs attached to each crystal.
The parameters of the pulses from each PMT were used to
apply various cuts to remove the noise events
from the analysis.
Our standard procedure of data analysis
involved fitting a single 
exponential to each integrated pulse in order 
to obtain the index of the exponential, 
$\tau$, the amplitude of the pulse, $A$, and the start time, $t_s$.
The scintillation pulses from nuclear and electron recoils were
shown to be nearly exponential in shape \cite{dan1}.
In addition to these, the mean time of the pulse (mean photoelectron 
arrival time), as an estimate
of the time constant, $\chi^2$ of the fit, number of photoelectrons and 
the energy were also calculated for each pulse. Conversion of the 
pulse amplitude to the number of photoelectrons and the energy was achieved
using pre-determined conversion factors, which come from
energy and single photoelectron calibrations. These calibrations
were performed typically every 2-4 weeks to ensure that there
were no changes in the PMT gains or crystal degradation.
Calibration with various gamma-ray sources provided 
the energy resolution of crystals \cite{naiad1}.

For each run (or set of runs) the ``energy -- time constant''
($E-\tau$) distribution was constructed. If all operational settings 
(including temperature) were the same for several runs, the
($E-\tau$) distributions for these runs were summed together.
Data collected at crystal temperatures outside a predefined range
were removed from the analysis. Each event 
stored on the computer disk, had a temperature
record in it. The mean daily temperature of each crystal
was plotted as a function of time, and the mean temperature
for a particular data set was calculated. The most common reason
for a temperature to fluctuate significantly (several degrees)
from the mean value was chiller failure. Data collected
without cooling were removed from the analysis. When a broken
chiller was replaced or repaired, it was impossible to achieve
the same crystal temperature. In this case we considered the two
data sets collected at different temperatures as independent and
did not combine the data but only the final limits \cite{naiad1}.

To reduce PMT noise and, particularly, events where
a spark (flash) in the dynode structure of one PMT
was seen by both PMTs, various cuts described in details in 
\cite{matt,noise} were applied. These cuts are an improvement
upon those used in Ref. \cite{naiad1}, having been shown to reject
spurious events more efficiently while retaining a larger fraction
of genuine scintillation pulses \cite{matt,noise}. The noise cuts
were developed based on the experiment with two PMTs mounted face to
face without the crystal between them. They were shown to cut similar
fraction of scintillation pulses from electron and nuclear recoils
(to within 2\%) \cite{matt,noise}. The cut efficiency for nuclear
recoil events from the surface calibration ranges from
84\% at 4-6~keV up to almost 100\% above 10~keV \cite{matt,noise}
for one of the crystals. As these cuts were found to be similar
(to within 2\%) for electron and nuclear recoils, the cut
efficiencies for each particular crystal and data set were determined
from the daily calibrations with the $^{60}$Co gamma source.
This also allowed the monitoring of the stability of these cuts.

For any small energy bin (1 keV width, for example), the time
constant distribution was approximated by a Gaussian in 
$\ln(\tau)$ ($\log$(Gauss) function) \cite{naiad1,ukdmc1,vak,dan2}
with three free parameters: mean time constant
$\tau_o$, width $w$ and normalisation factor $N_o$. 
In experiments where a second population
is seen (for example, nuclear recoils from a neutron 
source or possible WIMP-nucleus interactions), the resulting 
$\tau$-distribution was fitted with 
two $\log$(Gauss) functions with the same width $w$, since
the width is determined mainly by the number of collected 
photoelectrons. 

The aim of the analysis procedure was to find a
second population of events in the $\tau$-distributions or
to set an upper limit on its rate. To reach this,
the $\tau$-distributions for all energies of 
interest (2-40 keV) were obtained for all crystals
with gamma-ray ($^{60}$Co) and neutron ($^{252}$Cf) sources. 
Photons from high-energy
gamma-ray sources produce Compton electrons in the crystal
volume similar to those initiated by gamma-ray background in the
crystal. Neutrons collide elastically with the nuclei of the
crystal giving nuclear recoils similar to those expected
from WIMP-nucleus elastic scattering.
Calibrations were
performed on all crystals at the surface prior to moving the
crystals underground and were described in detail
in Refs. \cite{naiad1,vak}.

The crystal temperature and the light yield are the most important
parameters which affect the time constant distributions 
and, finally, the results of the experiment.
The temperature affects the pulse shape -- mean time of the pulse
and the index of an exponential. 
Increasing temperature causes the mean time (and the index of the fitted
exponential) to decrease at a
rate of about 5 ns per 1$^{\circ}$C. If the temperature of the crystal is
not stable, then combining data collected over a long period of time,
we will broaden the time constant distribution.
The same effect is observed if the light yield is not stable.
To improve the quality of the data, the crystal temperature was
controlled and monitored. To compensate for any small variations
of the crystal temperature and light yield over long periods of time 
(several months)
the gamma calibration of the crystal was done every day for 2 hours
with the $^{60}$Co gamma-ray source. In this way, any variation
in the mean time constant and width of the time constant distributions
of the real data, due to the effects mentioned above, is repeated in
similar distributions constructed with gamma calibration data and used
to fix some parameters in the fitting procedure (see Ref. \cite{naiad1}
for a full description of the data analysis procedure). 
Note that the ratio of mean time
constant for nuclear recoils to that of electron recoils does not 
depend on temperature \cite{naiad1} and was taken from the surface 
calibrations.

\vspace{0.5cm}
{\large \bf 4. Results and discussion}
\vspace{0.3cm}

The data from 6 crystals were used to set the limits
on WIMP-nucleus cross-section reported here. One crystal was excluded 
from the data analysis due to its small mass (4 kg) and high background
rate (about 15 events/kg/day/keV).
Table \ref{stat} shows the main characteristics and statistics for
all detectors for years 2002-2003. Statistics for the 10.6~kg$\times$years
of data collected in the period 2000-2001 are to be found
in Ref. \cite{naiad1}.

An energy range from 4 to 10 keV was used in the data
analysis. Previous results \cite{naiad1} were obtained for 
the energy range 4-30 keV. 
The smaller energy range in the present analysis was due to the smaller
amplitude range set on the 8-bit waveform digitisers. A total of 256 
digitisation points for a smaller amplitude range allowed a more accurate
calculation of the amplitude and time constant for single pulses
at low energies and better noise suppression.

\begin{figure}[htb]
\begin{center}
\epsfig{figure=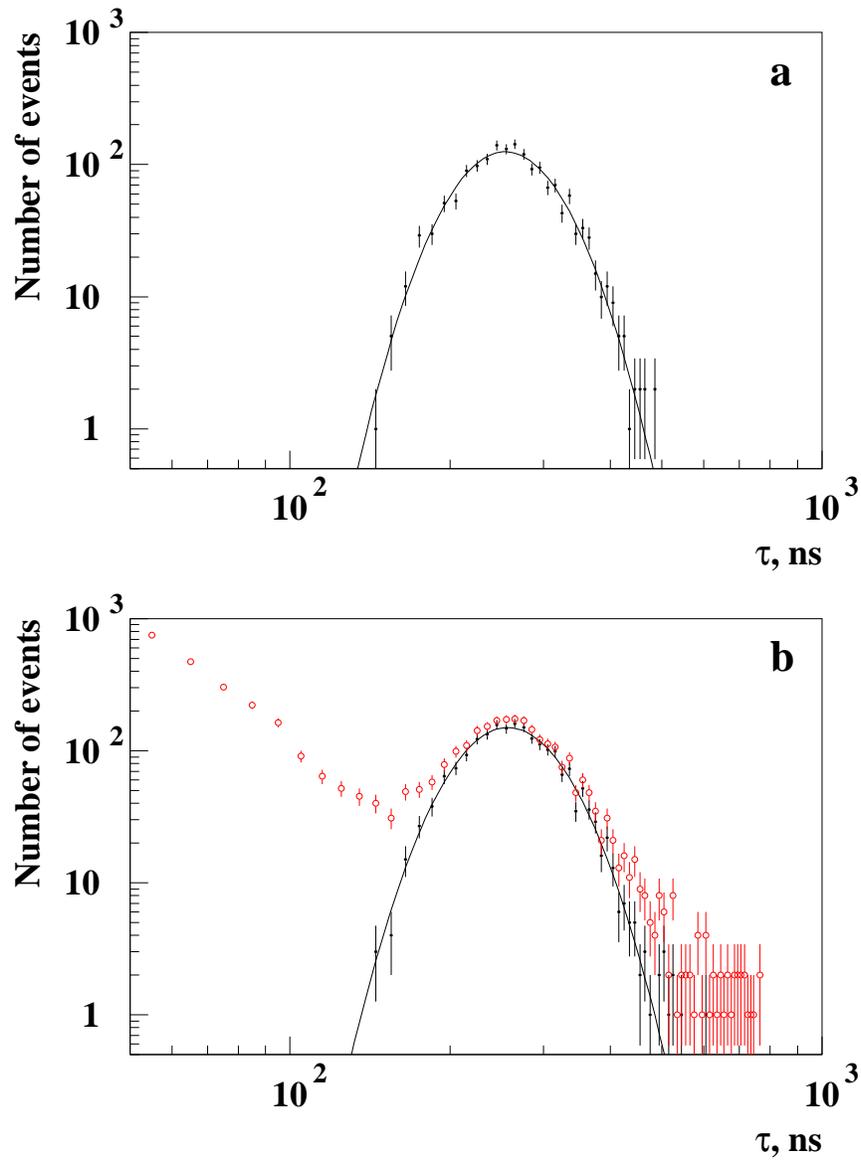,height=16cm}
\caption {Typical time constant distributions 
(shown here for crystal DM74) at energies 5-6 keV:
{\it a} -- for gamma calibration run (after cuts),
{\it b} -- for data before (open circles) and after (filled circles) 
cuts.}
\label{tcd}
\end{center}
\end{figure}

\begin{figure}[htb]
\begin{center}
\epsfig{figure=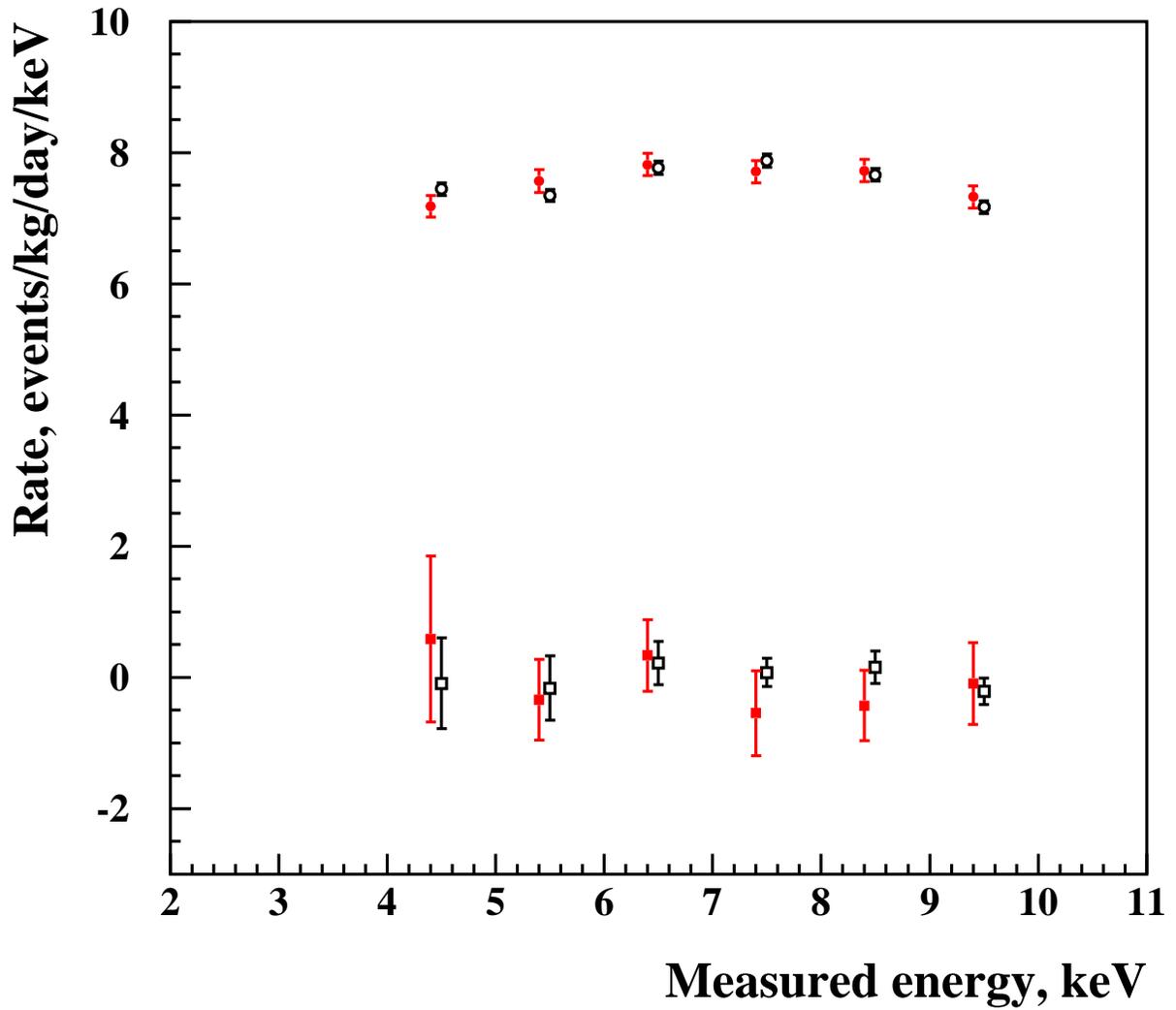,height=15cm}
\caption{Energy spectra after cuts from two runs of
one of the crystals (DM74) (filled and open circles)  
and the nuclear recoil rate for 
various energy bins (filled and open squares) with error bars 
drawn at 90\% C.L. Energy bins are: 4-5 keV, 5-6 keV, etc.
The points are shifted with respect to each other along x-axis
to avoid overlapping.}
\label{spectrum}
\end{center}
\end{figure}

Figure \ref{tcd} shows typical time constant
distributions at 5-6 keV from data (before and after cuts)
and a calibration run with gamma-ray source (after cuts).
PMT noise events are seen at small values of time constant before
cuts but are absent after the cuts were applied.
Figure \ref{tcd} does not reveal any visible difference
between data and calibration runs in terms of time constant
distributions.
Limits on the nuclear recoil rate at any particular energy
were obtained by fitting the measured time constant distribution
with two $\log$(Gauss) functions having known parameters: mean
time constants and widths, known from Compton and neutron 
calibrations. 
Free parameters were the total numbers
of electron and nuclear recoils 
(see Ref. \cite{naiad1} for further details). 
When both positive and negative values for the numbers of nuclear
and electron recoils were allowed, the best fit
values for nuclear recoil rates were either positive or negative but
normally within 1.5 standard deviations
from zero. 
Total event rates and nuclear recoil rates extracted from the
best two log-Gaussian fits to the data are shown in Figure \ref{spectrum} 
for two runs with crystal DM74. The error bars for nuclear recoil rates
correspond to 90\% C.L. As can be seen, in none of the energy bins
does the nuclear recoil rate deviate significantly from zero.
Hence no contribution from WIMP-nucleus interactions were observed
in these data.
This was found to be true for all crystals.
Since negative values are non-physical, to set upper limits on nuclear
recoil rates we restricted the rates to non-negative values and carried out
a two component fitting to the time constant distributions. To derive an
upper limit at 90\% C.L. we set the value of 
$\Delta (\chi^2)=\chi^2_{up} - \chi^2_{min} = 2.7$.
Strictly speaking this $\Delta (\chi^2)$ corresponds to the 90\% C.L.
only if the number of searched events is equal to 0. For
negative values of best fit number of nuclear recoils the value for an 
upper limit obtained in this analysis is higher than suggested in 
Ref. \cite{cousins} where a unified approach to confidence intervals was 
discussed for the Gaussian-with-boundary problem. 
For positive values of best fit number 
of nuclear recoils we used the number of recoils at $\chi^2_{up}$
as an estimate for an upper limit, which again gave us a higher value
than would follow from Ref. \cite{cousins}. Our approach was thus
conservative and allowed us to avoid complications with different
approaches suggested by different authors (see Ref. \cite{cousins} and
references therein for discussion).
In this way, the upper limits on the nuclear
recoil rate were obtained for each energy bin and for each
crystal. 

The limits obtained on the nuclear recoil rate for each energy bin and 
each crystal 
were converted into limits on the WIMP-nucleon spin-independent
and WIMP-proton spin-dependent cross-sections following the 
procedure described by Lewin and Smith \cite{ls} and used previously
in Ref. \cite{naiad1}. Expected
nuclear recoil spectra from WIMP-nucleus interactions were
calculated for a model with spherical isothermal dark matter
halo with parameters:
$\rho_{dm}$ = 0.3 GeV cm$^{-3}$, $v_o$ = 220 km/s, 
$v_{esc}$ = 600 km/s and $v_{Earth}$ = 232 km/s.
For the spin-independent case the form factors were computed
using Fermi nuclear density distribution with the parameters
to fit muon scattering data as described in Ref. \cite{ls}.
The WIMP-nucleus scattering cross-section was taken proportional
to $A^2$.
For the spin-dependent case a pure higgsino was assumed.
The spin factors and form factors were computed
for sodium and iodine nuclei on the
basis of nuclear shell model calculations \cite{ressell}
with Bonn A potential.
The quenching factors (scintillation efficiencies for nuclear recoils
with respect to that for electron recoils)
were taken as 0.275 for sodium and 0.086 for iodine
recoils \cite{dan1}.

The limits on the cross-section for various energy bins, targets
(sodium and iodine) and crystals were combined
following the procedure described in Ref. \cite{ls}
and used previously in Ref. \cite{naiad1},
assuming the measurements for different energy bins
and different crystals were statistically independent.

\begin{figure}[htb]
\begin{center}
\epsfig{figure=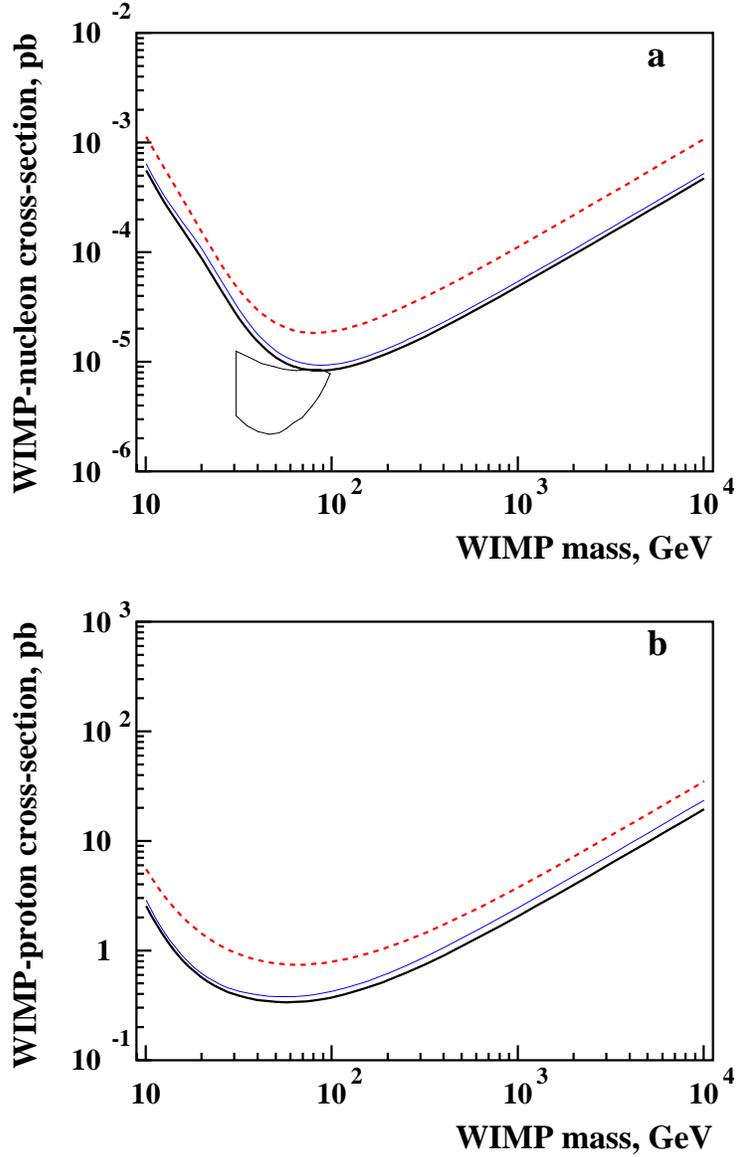,height=16cm}
\caption {NAIAD limits (90\% C.L.)
on WIMP-nucleon spin-independent (a) and WIMP-proton spin-dependent
(b) cross-sections as functions of WIMP mass:
dashed curves -- previously published limit
\cite{naiad1}, thin solid curves -- limit from 2002-2003 data,
thick solid curves -- combined limit (2000-2003).
Also shown is the 
region of parameter space favoured by the claimed DAMA positive
annual modulation signal (DAMA/NaI-1 through DAMA/NaI-4) 
(closed curve).}
\label{limits}
\end{center}
\end{figure}

Figure \ref{limits}a (\ref{limits}b) shows the NAIAD limits
on WIMP-nucleon spin-independent (WIMP-proton spin-dependent)
cross-sections as functions
of WIMP mass based on the data described in Table \ref{stat}.
Also shown in Figure \ref{limits}a is the 
region of parameter space favoured by the DAMA positive
annual modulation signal \cite{dama}.

\begin{figure}[htb]
\begin{center}
\epsfig{figure=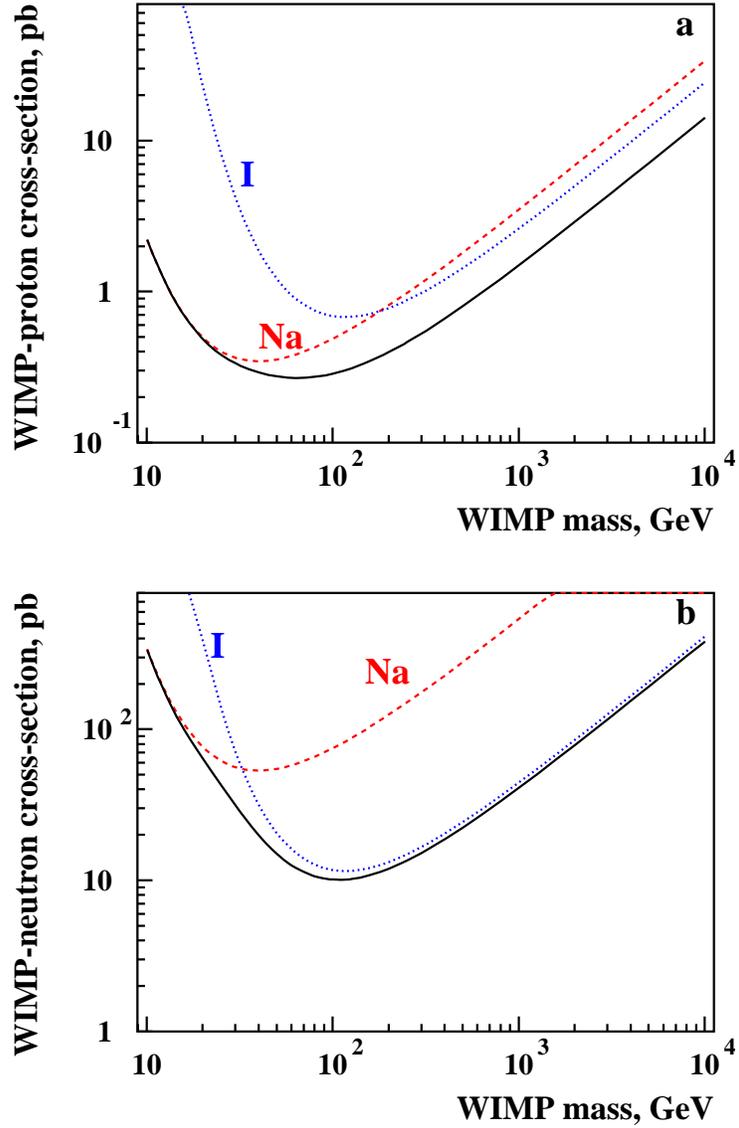,height=16cm}
\caption {Model-independent limits (90\% C.L.)
on spin-dependent WIMP-proton (a) and WIMP-neutron
(b) cross-sections as functions of WIMP mass.
The limits were derived following the procedure described
in Ref. \cite{dan3}. Dashed curves show the limits extracted from
interactions with sodium, dotted curves show those from iodine, and
solid curves show combined limits.}
\label{limits1}
\end{center}
\end{figure}

Model-independent limits on spin-dependent WIMP-proton and WIMP-neutron
cross-sections, calculated following the procedure described in 
Ref. \cite{dan3}, are presented in Figure \ref{limits1}. 
The NAIAD experiment imposes the most stringent constraints so far 
on the spin-dependent WIMP-proton cross-section (see
Refs.~\cite{edelweiss,giuliani} for a compilation of recent results).

\vspace{0.5cm}
{\large \bf 5. Conclusions}
\vspace{0.3cm}

Results from the NAIAD experiment for WIMP
dark matter search at Boulby mine were presented. 
Pulse shape analysis was used
to discriminate between nuclear recoils, possibly caused
by WIMP interactions, and electron recoils due to
gamma-ray background. We obtained upper limits on the 
WIMP-nucleon spin-independent and WIMP-proton spin-dependent
cross-sections based on the data accumulated by 6 modules 
(44.9 kg$\times$years exposure). The NAIAD experiment imposes the most 
stringent constraints so far on spin-dependent 
WIMP-proton cross-section.

\vspace{0.5cm}
{\large \bf 6. Acknowledgements}
\vspace{0.3cm}

\indent The Collaboration wishes to thank PPARC for financial support.
We are grateful to the staff of Cleveland Potash Ltd. 
for assistance.
We would also like to thank Prof. G.~Gerbier, Drs. J.~Mallet, 
L.~Mosca (Saclay), and Dr. C.~Tao (CPPM, Marseille) for providing us 
with one of their crystals for the NAIAD experiment.
C.~Duffy work was supported by the Nuffield Foundation through the
Undergraduate Research Bursary in Science. We also acknowledge
the funding from EU FP6 programme -- ILIAS.

\pagebreak

\begin{table}[htb]
\caption{Statistics for NAIAD detectors running during years 
2002-2003.}
\vspace{1cm}
\begin{center}
\begin{tabular}{|c|c|c|c|c|}\hline
Crystal & Mass, kg & Light yield, pe/keV & Time, days & Exposure, 
kg$\times$days \\
\hline
DM70-Saclay         & 10.0 & $5.1 \pm 0.4$ & 157.2 & 1571.7  \\
DM74                & 8.40 & $8.4 \pm 0.4$ & 298.3 & 2505.4  \\
DM76                & 8.32 & $4.6 \pm 0.3$ & 131.0 &  1089.6 \\
DM77                & 8.41 & $6.1 \pm 0.3$ &  221.9 &  1866.2 \\
DM80                & 8.47 & $8.0 \pm 0.5$ &  312.0 &  2642.7 \\
DM81                & 8.47 & $6.6 \pm 0.4$ &  336.2 &  2847.2 \\
\hline
Total exposure    &      &               &       & 12523  \\
\hline
\end{tabular}
\end{center}
\label{stat}
\end{table}
\pagebreak

\pagebreak

\pagebreak

\pagebreak

\end{document}